\documentclass[%
reprint,floatfix
amsmath,amssymb,
prb,
]{revtex4-2}

\usepackage{graphicx}
\usepackage{amsmath}
\usepackage{float}
\usepackage{braket}
\usepackage{dcolumn}
\usepackage{bm}

\begin{document}
	
	\preprint{APS/123-QED}

	\title{Topological Phase Transition in Commensurate \\ Multi-Frequency Floquet Su-Schrieffer-Heeger Model}
	
	\author{Sam Olin}
	\email{solin1@binghamton.edu}
	\author{Wei-Cheng Lee}%
	\email{wlee@binghamton.edu}
	\affiliation{%
		Department of Physics, Applied Physics, and Astronomy, Binghamton University, Binghamton, New York, 13902, USA
	}%
	
	\date{\today}
	
	\begin{abstract}
		Recently, Floquet systems have attracted a great deal of interest as they offer unprecedented ability to engineer topological states through the tuning of an external time-periodic drive. Consequentially, seeking new driving protocols that allow for more exotic topological phases and transitions becomes imperative for the Floquet engineer. In this paper, we study the Su-Schrieffer-Heeger model driven by two time-dependent periodic sources with commensurate frequencies and an amplitude modulation. Imposing more than one driving frequency allows us to realize even more exotic topological phases resulting from new couplings appearing in the Fourier space representation. Moreover, we find an experimentally practical method for sweeping the system through a topological phase transition by varying the amplitude mixture of the commensurate sources. We employ the local Chern marker, a real space representation of the Chern number, to simulate topological phase diagrams of the two-drive Floquet Hamiltonian in a variety of driving cases.
	\end{abstract}
	
	\maketitle
	\section{\label{sec:introduction}Introduction}
	Topology has signaled a shift in modern condensed matter research since the emergence of the quantum Hall effect \cite{PhysRevLett.45.494}. Many systems with desirable physical behaviors are now known to have an underlying nontrivial topological classification \cite{10.2307.2397741,PhysRevLett.49.405,PhysRevLett.61.2015,PhysRevB.22.2099}. These systems are promising platforms to potentially revolutionize technology as we approach the limit of traditional semi-conductor based devices \cite{ieee-roadmap}. One idea is to design electronics using the robust conducting edge states of topological insulators \cite{electronics7100225,2021CmPhy.4.70G} that could replace standard transistor-based switching components. Another idea is to leverage topologically protected spin-locked states as a basis for memory in spin-tronics \cite{2018NatPh.14.242S}. Finally, topological insulators and superconductors \cite{2012majorana} have been proposed as a basis for quantum computing \cite{osti_1811722}. With all of these possible applications at stake, it is clearly imperative that we maximize our ability to design and control topological phases of matter. 
	
	Consequently \textit{Floquet engineering} \cite{Okaengineering}, where systems are governed by a Hamiltonian possessing dynamic periodicity, has emerged as a promising candidate for precise tuning of topology through the laser-matter interaction. Floquet engineering has already been used to both emulate the Thouless pump with quantized energy as opposed to charge \cite{PhysRevLett.120.150601,Pan:21}, and as a method to create novel topological phases from initially trivial phases \cite{rudner2013anomalous,2011TNt-qwell,2011TNt-majorana,PhysRevB.79.081406,PhysRevB.84.235108,PhysRevLett.110.016802-mod-f-ti}.
	The recent growth in this field is owed to advances in experimental capabilities and theoretical understanding \cite{RevModPhys.53.287,1986PhR.141.320M,FAINSHTEIN1992111,CHU20041,doi:10.1142/5476} in the field of laser-driven quantum mechanics. One powerful theoretical tool in the field of driven lattice systems \cite{PhysRevLett.110.200403}, establishes a mapping of the dynamic D-dimensional system to a static D$+1$-dimensional system, in which the modes of the Fourier expansion play the role of lattice points in the new direction. An intriguing feature of the mapping is the emergence of a frequency-dependent field along the new frequency space direction - arising uniquely from the time derivative in the Schr\"{o}dinger equation. This field dictates the method of solution to be employed. In the case of adiabatic driving where the driving frequency is small, the frequency-dependent field is negligible or perturbative, and translational invariance along the frequency direction is assumed. For example, an adiabatically-driven $1$D Hamiltonian may be mapped to a $2$D static representation, where the Floquet-Bloch \cite{PhysRevLett.110.200403} formalism allows for standard calculations of the Chern number \cite{xiao2010,fukui_2005} to classify the topology. Alternatively in the high-frequency regime, couplings between neighboring Fourier modes become perturbative with the unperturbed Hamiltonian being time-derivative operator \cite{Eckardt_2015}. In the intermediate-frequency region, because the energy scales of the Floquet field and the static Hamiltonian are comparable, approaches beyond perturbation theory should be employed.
	
	In this paper, we study the topological properties of a system driven by two distinct frequencies in the adiabatic and intermediate frequency regimes. Floquet engineering is often employed for a single frequency drive, with multi-frequency (MF) cases being studied more recently. Broadly speaking, there are two options within the MF formalism: frequencies with (i) commensurate \cite{PhysRevA.101.032116} and (ii) incommensurate \cite{Martin_2017} relationships. The formalism of incommensurate multi-frequency driving demands the introduction of a Fourier manifold for each additional drive \cite{Martin_2017}, which has yielded useful application for $0$-dimensional qubit frequency mixers \cite{PhysRevX.12.021061}. However, this formalism and computation could be cumbersome for two dimensional systems and above. Moreover, the results in Floquet formalism rely on truncation of the typically infinite-dimensional Fourier mode space. Two competing truncation methods for each space of a two-tone incommensurate drive may yield results only applicable in limited situations. Commensurate driving on the other hand, has already been employed in a variety of situations with great experimental impact. Topology of $1$D lattice systems under commensurate driving has been studied before \cite{PhysRevB.102.235143} to examine the quality of localization in edge-modes. Another example is commensurate frequency driving being used to create quantum destructive interference in a Fermi-Hubbard model to suppress heating effects \cite{PhysRevX.11.011057}, which is a prevalent problem in all of Floquet engineering. Finally two-tone drives have been used to engineer non-trivial band structures \cite{Sandholzer_2022,Minguzzi_2022}. 
	
	We employ the commensurate frequency framework developed in \cite{PhysRevB.102.235143} to express the MF Floquet formalism using a single period, and apply this drive to The Su-Schreiffer-Heeger (SSH) model. Note that single frequency driven SSH variations have already generated great interest in the field \cite{PhysRevA.92.023624,Borja_2022,dmytruk2022controlling,Agrawal_2022}. In the adiabatic driving scheme, we map the commensurate drives to the frequency space, resulting in the emergence of new couplings. Careful tuning of the frequencies would allow for simulating those nontrivial couplings that are difficult to realize in the real-space. To demonstrate this, we explore effects such as a next-nearest neighbor Floquet hopping, and large-lattice hopping with two larger, close by frequencies. The latter effect motivated our interest in this study, as a potential temporal analogue to the Moir\'e pattern observed in twisted bi-layer graphene \cite{doi:10.1073/pnas.1108174108}. Finally, we demonstrate that the dual frequency drive provides an experimentally appealing method for creating a topological phase transition. The different topological regions are simply reached through varying the amplitude mixture.
	
	In simulating the model, we find that standard computational approaches to topology can be troublesome in the presence of the Floquet field originating in the intermediate frequency regime. We address this by employing a real space variant of the Chern number, called the local Chern marker \cite{2011bianco}. This method allows us to not only visualize the effect of open-boundaries, disorder, and electric field on the topology locally, but to also compute the system topology in the presence of a non-zero Floquet field, which is a feature completely unique to the Floquet engineering systems.
	\maketitle
	\section{\label{sec:MMFT}Model}
	A static Su-Schreiffer-Heeger (SSH) model is known to possess a dimer type lattice with atoms $A,B$ forming the members of each dimer, and it is topologically non-trivial provided that the inter-cell coupling is stronger than the intra-cell coupling. To demonstrate the consequences of multi-frequency driving, we consider a Floquet Su-Schreiffer-Heeger (FSSH) model with time varying, two-frequency tunneling coefficients. The Hamiltonian is kept similar to previous works \cite{PhysRevB.95.205125,L__2019} to ensure that upon relaxing the two-drive condition to a single drive, we recover well-established results.
	\begin{equation}\label{eq:FSSH}
		\begin{split}
			H(t)=\sum_{n}^N U_1(t) \hat{c}^\dagger_{n,B}(t)\hat{c}_{n,A}(t)+U_2(t) \hat{c}^\dagger_{n+1,A}(t)\hat{c}_{n,B}(t) \\
			V_A(t) \hat{c}^\dagger_{n+1,A}(t)\hat{c}_{n,A}(t) + V_B(t) \hat{c}^\dagger_{n+1,B}(t)\hat{c}_{n,B}(t) + h.c.
		\end{split}
	\end{equation}
	In Eq. \ref{eq:FSSH}, $U_1(t),U_2(t)$ are the intra-cell and inter-cell tunneling strengths, respectively, which are periodic in $T$. Additionally, we consider the next-nearest neighbor coupling terms $V_A(t),V_B(t)$. The real-space coordinate can have a periodic boundary condition ($N+1=1$) or an open boundary. The tunneling coefficients;
	\begin{equation*}
		\begin{split}
			U_1(t)=u(1+2(\cos{\Omega_1 t}+\alpha\cos{\Omega_2 t})) \\
			U_2(t)=u(1-2(\cos{\Omega_2 t}+\alpha\cos{\Omega_2 t})) \\
			V_A(t)=v(\cos{(\Omega_1 t+\theta)}+\alpha\cos{(\Omega_2 t+\theta)}) \\
			V_B(t)=v(\cos{(\Omega_1 t-\theta)}+\alpha\cos{(\Omega_2 t-\theta)}) 
		\end{split}
	\end{equation*}
	are dynamical with driving frequencies $\Omega_{1,2}$ and tunneling amplitudes $u,v$ for the nearest and next-nearest neighbor hopping, respectively. The $\Omega_2$ driving factor possess an "offset" amplitude $\alpha$, the consequences of which will be discussed in the Results section.
	\maketitle
	\subsection{Dual Frequency Driving}
	\begin{figure}[t]
		\centering
		\includegraphics[scale=0.375]{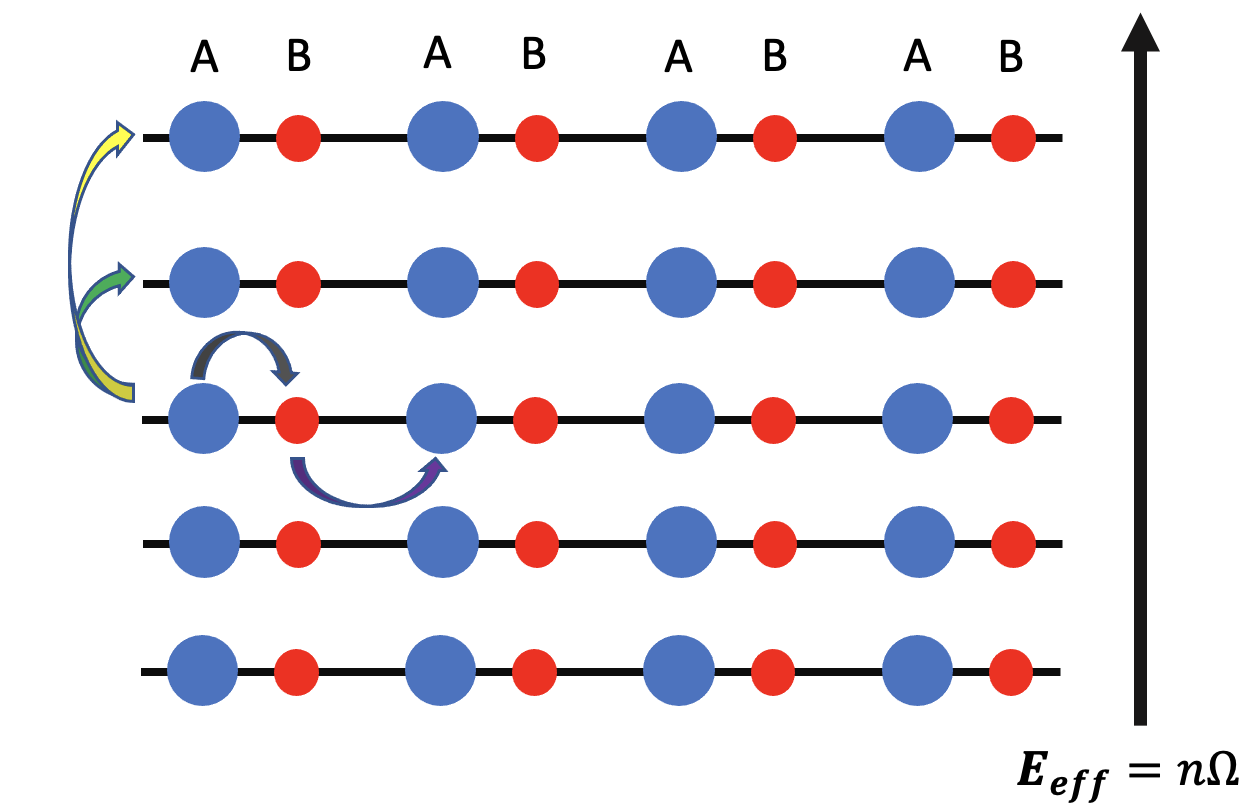}
		\caption[Chern Marker Visualized]{$2$D Static Floquet SSH representation. In gray: intra-cell hopping, purple: inter-cell hopping, green: nearest neighbor Floquet (NNF) hopping, yellow: next-nearest neighbor Floquet (NNNF) coupling. The basic couplings are shown but the Hamiltonian allows for any combination, e.g. NNNF+inter-cell off-diagonal element is present in $H$. Note NNNF couplings may only occur with the additional frequency $n_2=2$ drive.}
		\label{fig:fssh-2p}
	\end{figure}
	Here we outline the treatment of Eq. \ref{eq:FSSH} in which the sources are periodic in $T_1$ and $T_2$ and are subject to the following condition:
	\begin{equation}\label{ratio}
		\frac{T_1}{T_2}=\frac{n_2}{n_1}
	\end{equation}
	for $\{n_1,n_2\}$ $\epsilon$ $\mathbb{Z}^+$ meaning that we may always find \cite{PhysRevB.102.235143} a period, $T$, such that
	\begin{equation}
		T=n_1T_1=n_2T_2
	\end{equation}
	which may be used to employ the Floquet theory. Note that $\Omega_{1,2}=\frac{2\pi}{T_{1,2}}$. The Hamiltonian may be expressed in terms of "components" for each period,
	\begin{equation}\label{eq:H-comp}
		H(\textbf{r},t)=H_0+H^{T_1}(\textbf{r},t)+H^{T_2}(\textbf{r},t)
	\end{equation}
	where the components have the following periodicity: $H^{T_1 (T_2)}(\textbf{r},t+T_1 (+T_2))=H^{T_1 (T_2)}(\textbf{r},t)$, and the system has $H^{T_1 (T_2)}(\textbf{r},t+T)=H^{T_1 (T_2)}(\textbf{r},t)$. The $H_0$ term is the undriven Hamiltonian.
	It is important to note that the period of the system is $T$ and so Floquet theory is employed on $T$, not on either $T_1$ or $T_2$. Due to this condition, the derivation of the Floquet Hamiltonian, $\mathcal{K}$, is the same as in the single frequency case. The single-frequency Floquet Hamiltonian has been derived many times so we refer the reader to Refs. \cite{Okaengineering,https://doi.org/10.48550/arxiv.2003.08252}. The general procedure is as follows: from the time-dependent Schr\"{o}dinger equation, express the eigenstates as Floquet states that are composed of a non-periodic phase factor and a $T-$periodic function, then expand the states using a Fourier expansion. The single frequency case diverges from the dual frequency case as we take the expansion in terms of the components of Eq. \ref{eq:H-comp}. The components are written in Eq. \ref{mmode_k}.
	\begin{equation}\label{mmode_k}
		\mathcal{K}=\left( H^{T_1}(\textbf{r},t)+H^{T_2}(\textbf{r},t) \right)-i\frac{\partial}{\partial t}
	\end{equation}
	The Fourier expansion on the Floquet modes is then,
	\begin{sloppypar}
		\begin{equation*}
			\begin{split}
				\left( H^{T_1}(\textbf{r},t)+H^{T_2}(\textbf{r},t) \right) \sum_m e^{im\Omega t}\ket{\phi^m_\alpha}\\+\sum_m m\Omega e^{im\Omega t}\ket{\phi^m_\alpha}
				=\epsilon_\alpha \sum_m e^{im\Omega t}\ket{\phi^m_\alpha}
			\end{split}
		\end{equation*}
	\end{sloppypar}
	meaning that the matrix elements of Eq. \ref{mmode_k}, given by the universal equation: $\bra{\alpha,n}\dots\ket{\beta,m}=1/T\int_0^T dt \dots$ may computed as in the usual way. However, we must pay consideration to each new hopping term emerging from each commensurate frequency;
	\begin{equation*}
		\begin{split}
			\sum_{m,n}\int_0^T \textrm{d}t H^{T_1}(\textbf{r},t)e^{i\Omega t(m-n)}+\\ \int_0^T\textrm{d}t H^{T_2}(\textbf{r},t)e^{i\Omega t(m-n)} =\epsilon_\alpha -m\Omega \delta_{mn}
		\end{split}
	\end{equation*}
	resulting in
	\begin{equation}\label{eq:mmode_h}
		\sum_{m,n} \left(  H^{T_1}_{(m-n)}+H^{T_2}_{(m-n)} \right) \ket{\phi^m_\alpha}+m\Omega \delta_{m,n}\ket{\phi^m_\alpha}=\epsilon_\alpha \ket{\phi^n_\alpha}
	\end{equation}
	where $H^{T_1 (T_2)}_{(m-n)}=\frac{1}{T}\int_0^T \textrm{d}t H^{T_1 (T_2)}(\textbf{r},t)e^{i\Omega t(m-n)}$. Note that the Fourier factor $e^{i\Omega t (m-n)}$ is left in $\Omega$ the frequency of the system, not in either $\Omega_1$ or $\Omega_2$. The result of mapping our Hamiltonian to the static $2$D version is shown in Fig. \ref{fig:fssh-2p}. The new couplings seen there emerge from new frequencies added in the drive.
	The coupling factor in the Floquet Hamiltonian in Eq. \ref{eq:mmode_h} compared to the single-frequency case reveals that multi-mode theory with commensurate drives allows for construction of new kinetic terms. One may expect a new coupling for each commensurate drive added. Careful construction of these new frequencies may yield exotic new physics, or open the route for Floquet systems to mimic the physics of some experimentally intractable static systems in condensed matter. 
	
	\maketitle
	\section{\label{sec:Results}Results}
	The Hamiltonian (Eq. \ref{eq:FSSH}) is constructed for atoms $A,B$ with $20$ dimers, resulting in $40$ real space matrix elements. We set $u=1,v=0.2,\theta=0.5\pi,\alpha=2$, unless otherwise states. We employ the multi-frequency Floquet theory (Eq. \ref{eq:mmode_h}) to map the time dependent $1$D system to the static $2$D enlarged space (Fig. \ref{fig:fssh-2p}). As opposed to mapping the system via the Floquet-Bloch transformation, we leave the system in the real-space matrix form. While this matrix is technically infinite, we can study a truncated space using the Chern marker to examine the topological order \cite{https://doi.org/10.48550/arxiv.2003.08252,Holthaus_2015}. We consider $200$ Floquet modes, $m$, resulting in a $8000\times8000$ matrix unless otherwise stated, which we construct and diagonalize in Fortran. All presented calculations of the Chern number are accurate up to a maximum error of $1\%$. Where stated, the Stark field is considered by adding in the $m\Omega$ dependent value along the Floquet diagonal $\delta_{nm}$. As for the coupling, the integers chosen in Eq. \ref{ratio} result in different delta functionals after integration of Eq. \ref{eq:mmode_h} in the Floquet space due to the cosine drive. For example $n_1=10\rightarrow \delta_{n,m+10}+\delta_{n,m-10}$. Although theoretically any integer ratio may be employed, here we consider the $\{n_1,n_2\}$ cases of $\{1,1\}$ (single drive reference), $\{1,2\}$ (Floquet next-nearest neighbor), $\{10,11\}$ (close-by beat frequency).
	\subsection{Topology - Single Drive}
\begin{figure}[t]
		\centering
		\includegraphics[scale=0.40]{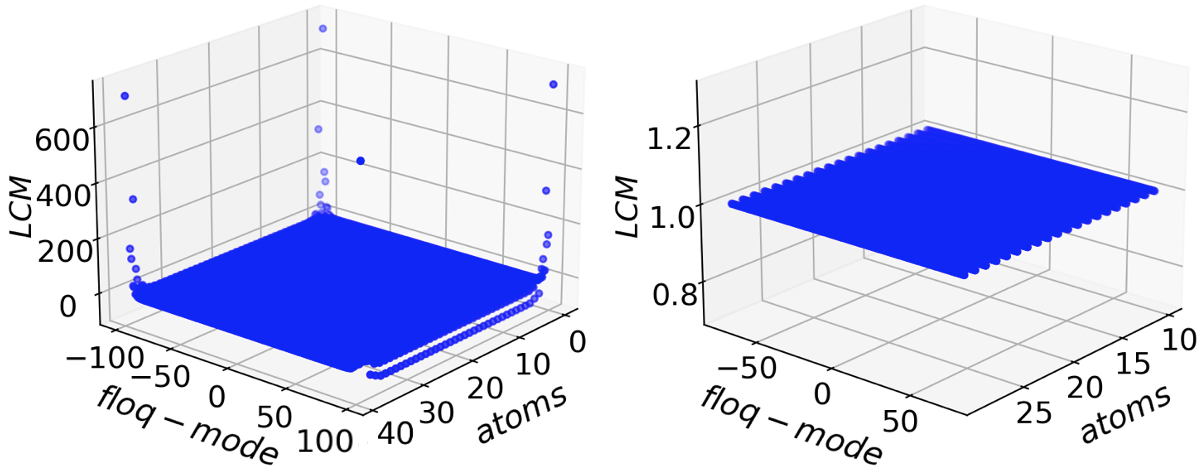}
		\caption[Chern Marker Visualized]{Two plots of the Chern marker. Left is the Chern marker over the whole system, in two dimensions, real space and Floquet space. The average Chern number is $0$, due to the trace identity \cite{2011bianco}. In the region away from the edge as shown in the right plot, we obtain a Chern marker of $1$, corresponding to the bulk topology. The marker is stable in the bulk, and non-physical near the edges.}
		\label{fig:simple chern marker}
	\end{figure}

\begin{figure}[t]
		\centering
		\includegraphics[scale=0.34]{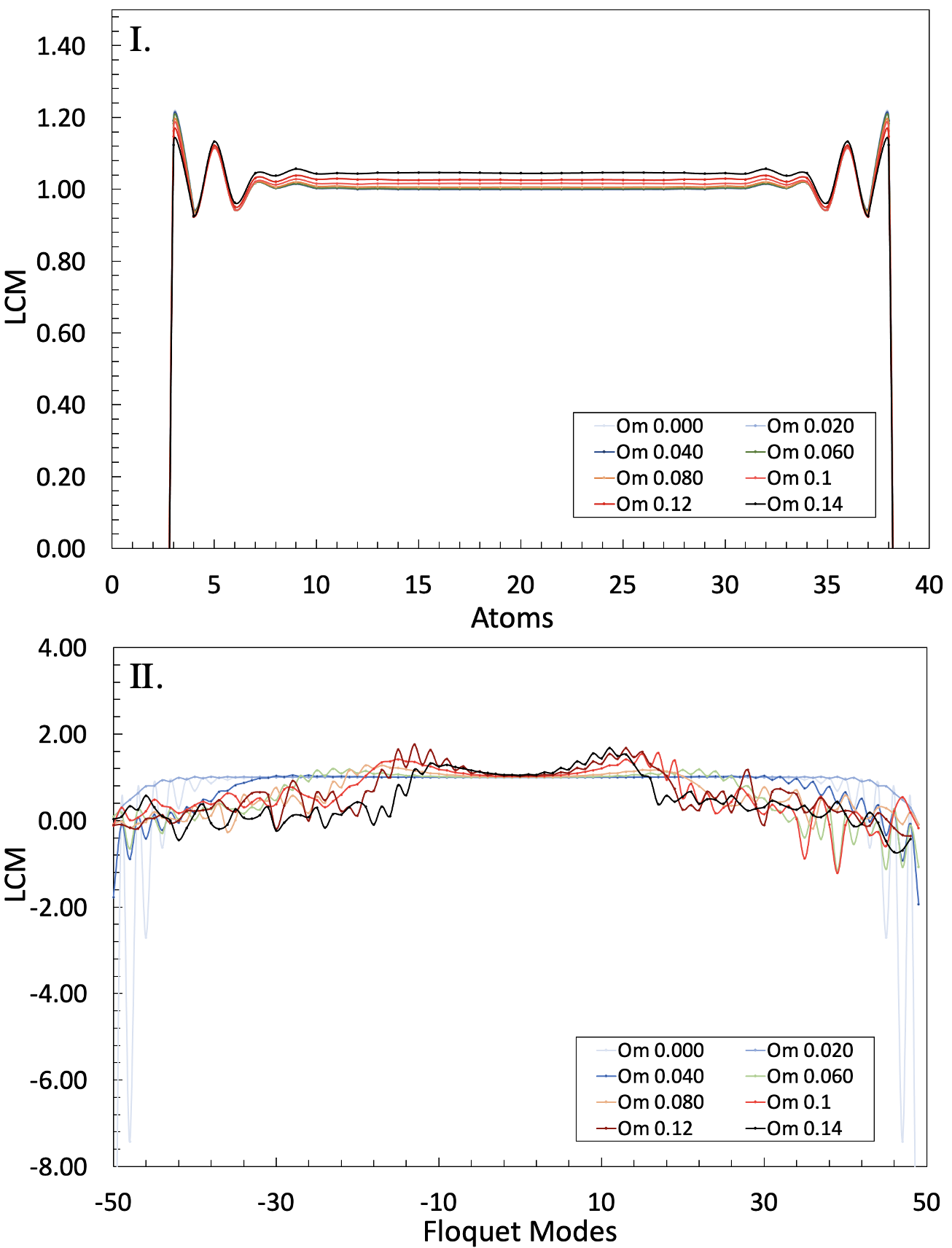}
		\caption{Chern marker plotted in I. real space and II. Floquet space, for increasing $\Omega$, magnitude of the Stark field. Note that only $100$ Floquet modes were used. In I. the real space Chern marker is not strongly changed by increasing $\Omega$ - a sensible result as the Stark field only permeates the Floquet space. In II. the Stark field destroys the topological order in the sample as $\Omega$ increases, but notice that for small $m$ the Chern marker returns to the expected Chern number of $1$, confirming the adiabatic theorem for small $\Omega$.}
		\label{fig:ch_vs_omega}
	\end{figure}

	Our model relaxes to a single drive case by setting $\alpha=0$, and setting $\Omega_1$ as the base frequency. In Fig. \ref{fig:simple chern marker}, we plot the Chern marker over the static $2$D representation of the sample in the presence of periodic boundaries. As expected \cite{2011bianco}, averaged over the entire sample the marker is $0$ due to its commutator definition. However, in the bulk of the sample, the average Chern marker yields $1$, in excellent agreement with the Fukui method. Previous works \cite{L__2019} have discovered that the single drive Hamiltonian is topologically non-trivial for non-zero $\theta$. However, these predictions enforce translationally invariant samples and rely on computational methods using $k$-space, meaning the effect of the Floquet field on the topological order is ignored. By employing the local Chern marker \cite{2011bianco} we provide both real-space confirmation of single drive topology, and simple determination of topology in the face of the Floquet electric field.  
	An advantage of computing topology using the Chern marker is that we may easily reintroduce the Floquet field along the frequency direction for small values of $\Omega$. Consequently, the topology may be visualized along each direction in response to increasing field value, or even disorder along the real-space direction. In Fig. \ref{fig:ch_vs_omega}, we plot the Chern marker along real and frequency space with increasing electric field $\propto \Omega$. We see the real space Chern marker unaffected everywhere by increasing $\Omega$. Similarly, for finite but small $\Omega$ the frequency space Chern marker remains unaffected. However as $\Omega$ increases the LCM along the frequency space does not remain topologically invariant and the system does not have a meaningful topology. At the very center of the sample where $m=n\approx 0$, the expected topology is recovered which is consistent with the adiabatic theorem for small fields $(m\Omega)$. Consequently, this calculation may be used to probe maximum allowed values of $\Omega$ above which the topology becomes ill-defined.
	\subsection{Multi-Frequency Drive: Case 1,2}\label{sec:1_2}
	\subsubsection{Topological Phase Transition}
\begin{figure}[t]
		\centering
		\includegraphics[scale=0.325]{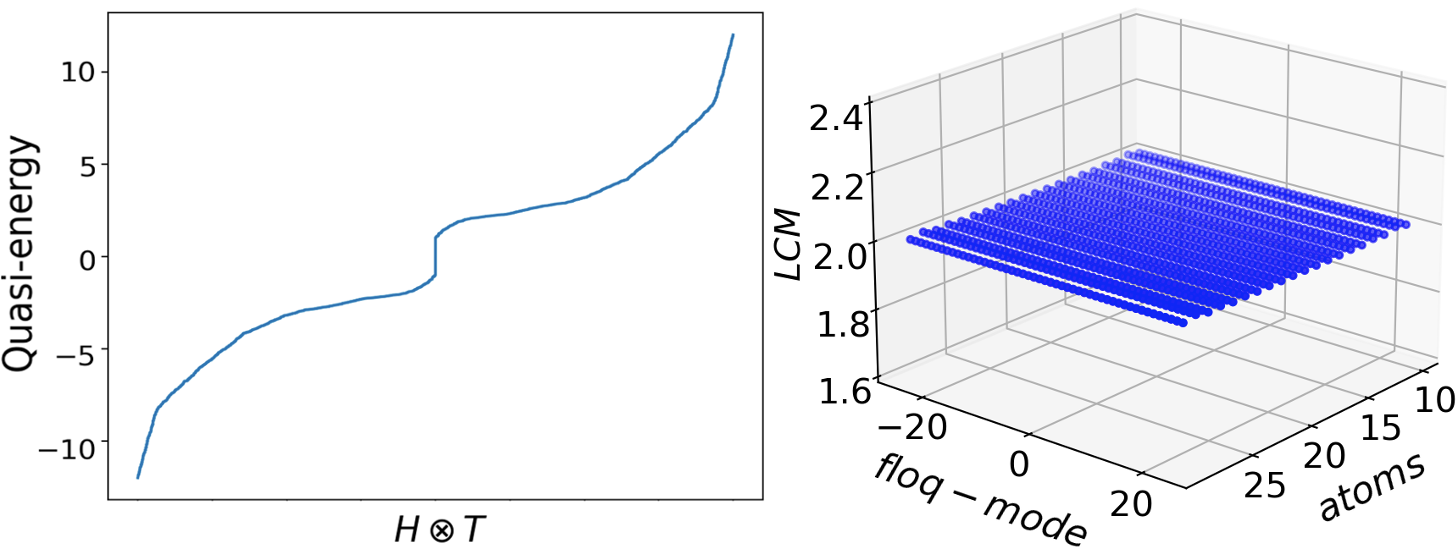}
		\caption[Next-Nearest Floquet Hopping]{The band (left) and Chern marker interior (right) for the case $n_1=1,n_2=2$. This case simulates a second neighbor hopping in the Floquet space - impossible to achieve without a second drive. The system is found to be insulating with a quasi-energy gap of $\sim 2.0$, meaning topology can be computed. The right plot shows the interior of the Chern marker, and the bulk average value of $2$. }
		\label{fig:NNN-floq}
	\end{figure}

\begin{figure}[t]
		\centering
		\includegraphics[scale=0.42]{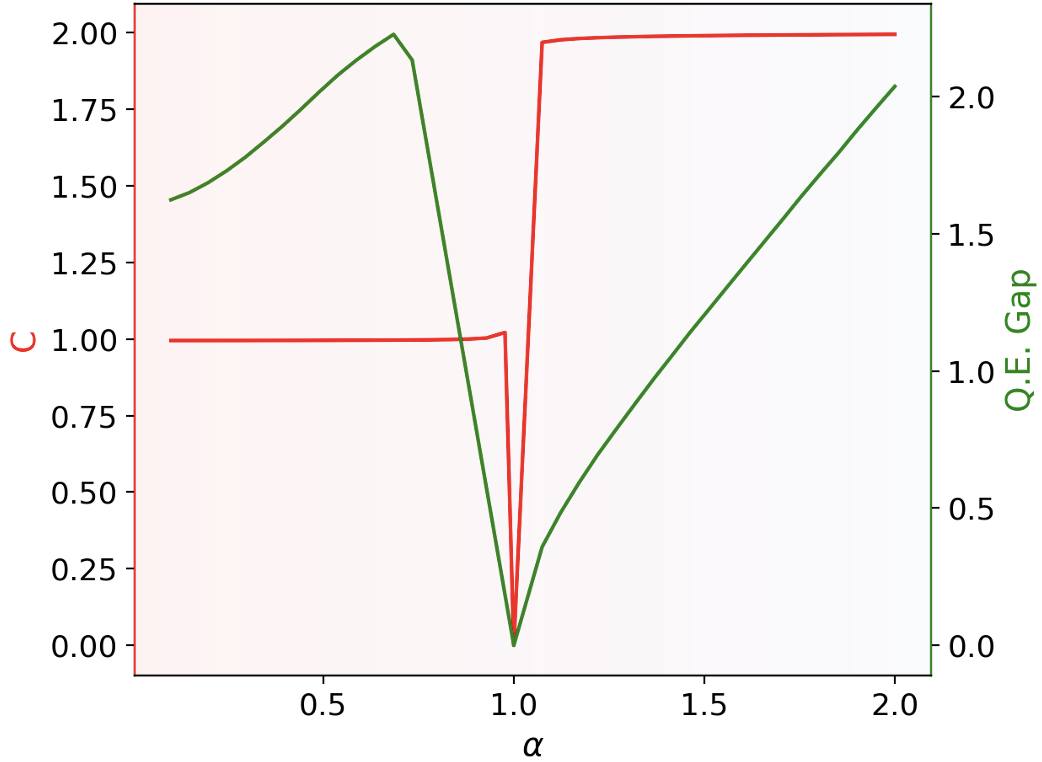}
		\caption[C vs Alpha]{The Chern number in red plotted against the $\Omega_2$ driving amplitude $\alpha$. Note that the amplitude on $\Omega_1$ is $1$. Below the transition point $1$, the Chern number is robustly $1$. Above the transition point the Chern number is robustly $2$. The quasi-energy gap is shown to close at the transition point in green. The color gradient is shown to signify that even changing parameters such as $\theta,u,v$ will yield the same results provided that the gap is not closed. }
		\label{fig:C-vs-alpha12}
	\end{figure}

	Here we examine the effects of the second drive in the case of frequency ratio $n_1,n_2=1,2$. This frequency ratio has been studied before for a variety of systems, \cite{Sandholzer_2022,Minguzzi_2022,PhysRevX.11.011057}, but to our knowledge, not in the driven SSH model. The band structure and the Chern marker for the case of periodic boundaries in each direction (hence adiabatic driving) are computed and plotted in Fig. \ref{fig:NNN-floq}. The physical interpretation of this model is the presence of a next-nearest neighbor coupling along the frequency space. The system still possesses a gap for $n_1,n_2=1,2$, but only the case that the amplitudes of the respective drives are different. Here we fix $\alpha=2$. Due to the invariance the Chern marker is seen to be relatively stable with little variation over the sample bulk.
	We find that the Chern number is $2$, indicating that an advantage of commensurate driving is the ability to engineer non-trivial phases with $C>1$. Surprisingly, we discover that the offset amplitude between the drives functions as a tuning parameter for a topological phase transition (TI). This transition is plotted in Fig. \ref{fig:C-vs-alpha12}. For $\alpha<1$, $C=1$, and for $\alpha>1$, $C=2$. The transition occurs through the gap closing condition of $\alpha=1$. Note that the gap is computed as the difference between the lowest conduction band and highest valence band quasi-energies. The gap closes on Brillouin zone corners $\{k_x,k_f\}=\{-\pi,-\pi\}$, etc., but initially the smallest difference is elsewhere in the Brillouin zone.
	This transition is not unique to the case of $n_1,n_2=1,2$. We expect this behavior for any choice of $\Omega_1,\Omega_2$ provided that the system remains gapped. The Chern number will transition with the amplitude mixture controlling the critical point, and the frequencies controlling the topology. This amplitude modulation of a two-frequency drive should be experimental feasible. It does not rely on fabrication of the lattice or on a quantum well thickness \cite{Bernevig_2006}. The formalism presented here poses an easily tune-able topological phase transition, based simply on the control of the driving lasers. However, it is also known \cite{Minguzzi_2022} that the relative phase of the two-frequency drive can change the symmetry of the system which is being examined in ongoing calculations.
	\subsubsection{Edge States}
\begin{figure}[t]
		\centering
		\includegraphics[scale=0.29]{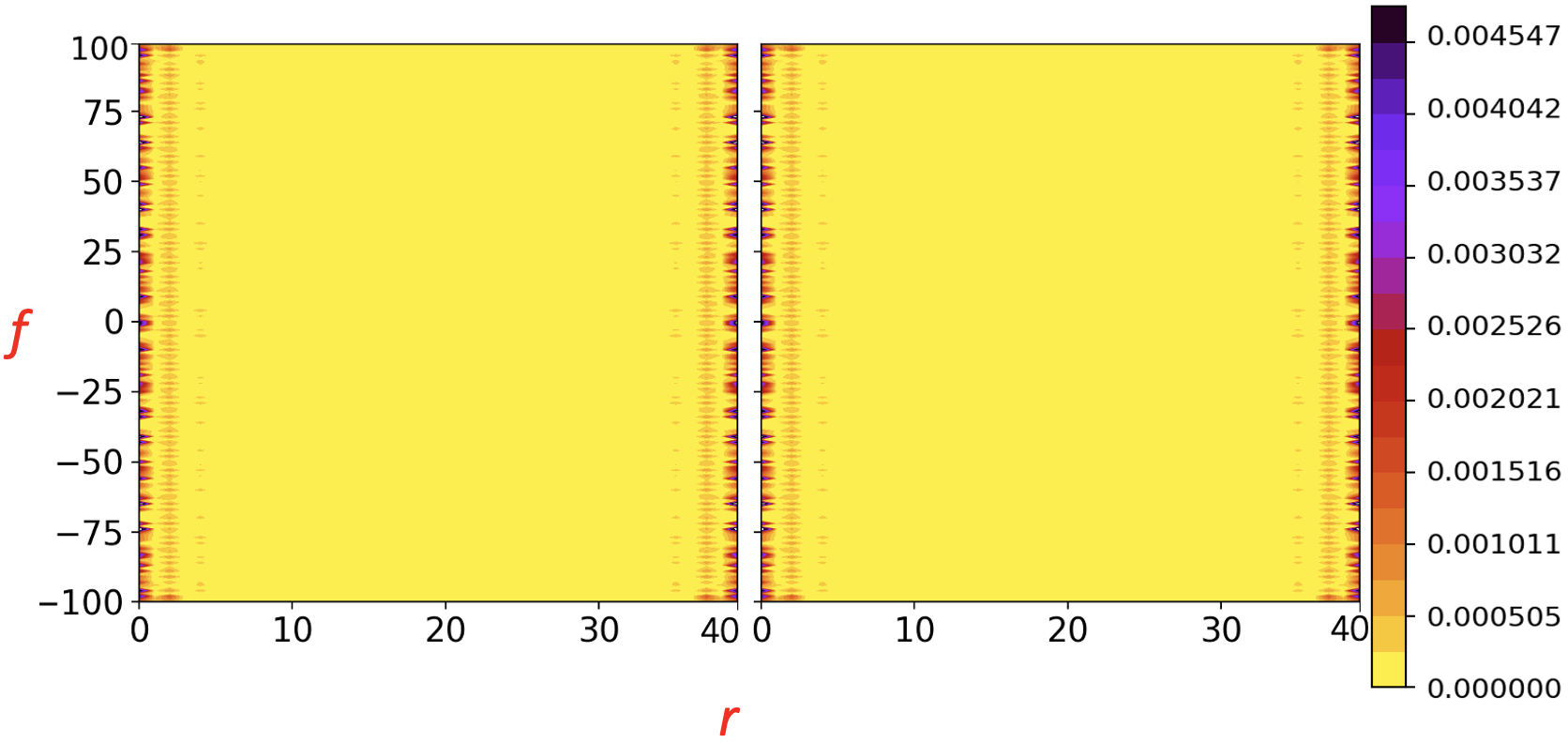}
		\caption[n-1-2]{Edge modes occurring along the real space edge for the case of $n_1,n_2=1,2$ driving frequency ratio. The contour is computed for $\psi^{\dagger}\psi_{i,j}$, for $i,j$ the real and frequency space, respectively. The states plotted are the $0$-quasi-energy states.}
		\label{fig:n-1-2}
	\end{figure}

	The topology present in this drive case has an observable physical effect, manifested in the emergence of edge modes along the edges of the effective $2$D sample. In Fig. \ref{fig:n-1-2} we plot the zero-energy eigenstates upon opening both the Floquet, and the real space boundary. The states are seen in Fig. \ref{fig:n-1-2} to occupy the real space boundary, with amplitude diminishing to $0$ in the center of the sample. This plot is constructed by taking $\psi^{\dagger}\psi_{i,j}$ for $i,j$ the real and frequency space elements, respectively, for the eigenstates at $0$-quasi-energy.
	The real-space treatment easily allows us to plot the zero-energy states for a variety of boundary conditions. For example one may reinstate translational invariance along the SSH chain, break invariance of the Fourier manifold, and recompute the amplitude of \ref{fig:n-1-2}. In this case, we find that the edge states exist along the "Fourier edge" only. Unlike SSH chain edge states which may be observable in current measurements, Fourier edges are only an artifact of the theoretical Floquet mapping and subsequent truncation scheme, so we neglect the result in the current paper.
	\subsection{Multi-Frequency Drive: Case 10,11}
	\subsubsection{Exotic Topology}
\begin{figure}[t]
		\centering
		\includegraphics[scale=0.35]{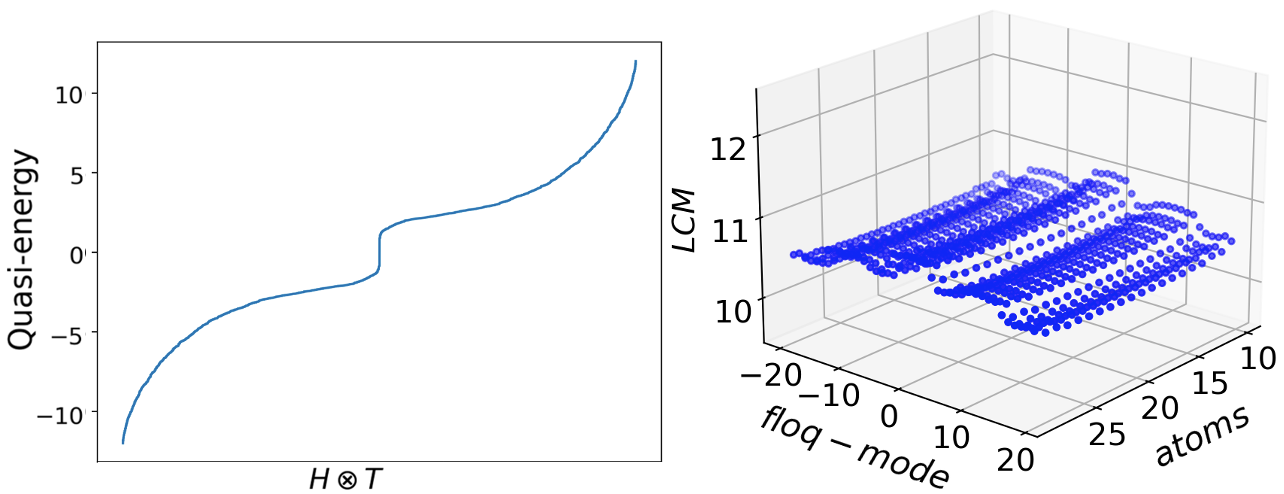}
		\caption[Long Range Floquet Hopping]{The band and Chern marker interior for the case $n_1=10,n_2=11$. The band is shown in the left, and the system is found to be gapped in quasi-energy, meaning topology can be computed. The right plot shows the interior of the Chern marker, and the bulk average value of $11$. Note that the $n_1,n_2=10,11$ shows more interference in the LCM than the $1,2$ case.}
		\label{fig:d-10-11}
	\end{figure}

\begin{figure}[t]
		\centering
		\includegraphics[scale=0.42]{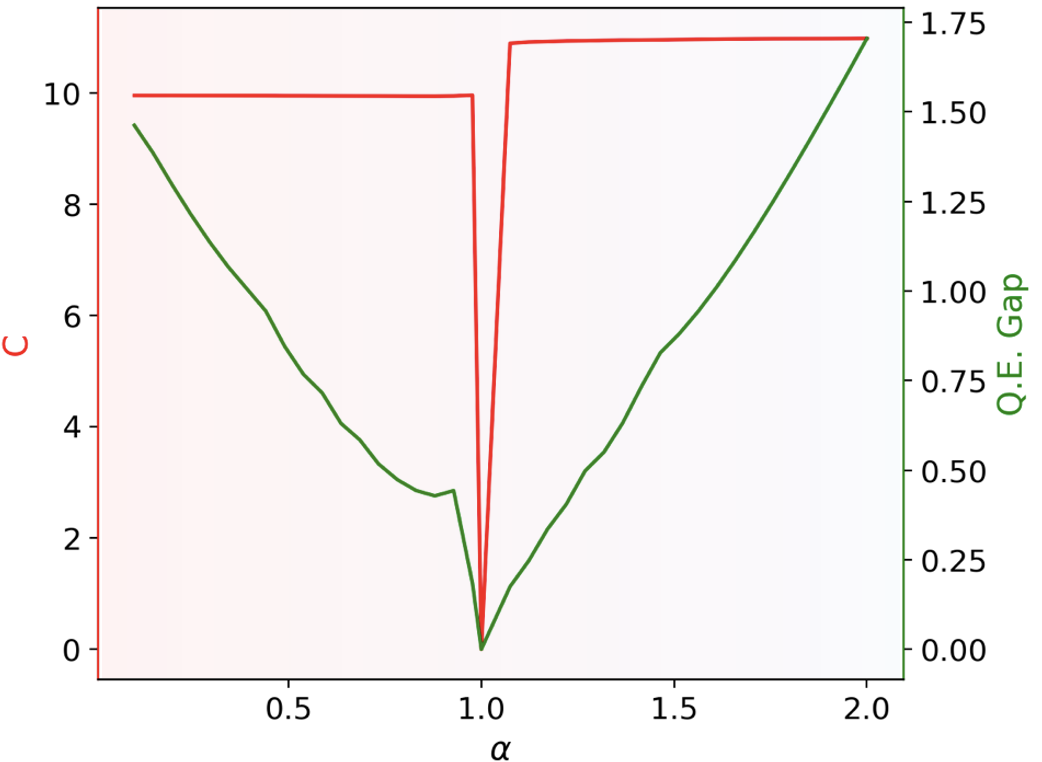}
		\caption[C vs Alpha]{The Chern number in red plotted against the $\Omega_2$ driving amplitude $\alpha$. Note that the amplitude on $\Omega_1$ is $1$. Below the transition point $1$, the Chern number is robustly $10$. Above the transition point the Chern number is robustly $11$. The quasi-energy gap closes at the transition point, of $\alpha=1$. The color gradient is shown to signify that even changing parameters such as $\theta,u,v$ will yield the same results provided that the gap is not closed.}
		\label{fig:C-vs-alpha1011}
	\end{figure}

	Here we examine the effects of the second drive in the case of frequency ratio $n_1,n_2=10,11$. The band structure and the Chern marker for the case of periodic boundaries in each direction (hence adiabatic driving) are computed and plotted in Fig. \ref{fig:d-10-11}. The system  possesses a gap for $n_1,n_2=10,11$, with condition $\alpha=2$. The Chern marker is seen to be not as stable over the sample bulk as in the $n_1,n_2=1,2$ case, resulting from interference between the two close-by frequencies. We again tune $\alpha$ through the critical point, as plotted in Fig. \ref{fig:C-vs-alpha1011}. The system displays the same phase transition behavior as in the $n_1,n_2=1,2$ case. The Chern number is found to be $C=10$ for $\alpha<1$ and $C=11$ for $\alpha>1$.
	Our motivation in studying the multi-frequency driving was to model a beat frequency Hamiltonian. This is based on the hope that the disorder induced by the beat frequency along the Floquet direction would be a temporal analog to the twisted bi-layer graphene, in which maximum disorder occurs for certain "magic" angles. The requirement then, is that the two drives possess frequencies which are close in value. The size of the matrix must accommodate long-range couplings. The case for $n_1,n_2=10,11$ is explored using our current Fortran code, but larger frequency ratios like $100,101$ demand a much larger matrix. This larger case would be ideal to consider for the beat frequency analogue.
	\subsubsection{Edge States}
\begin{figure}[t]
		\centering
		\includegraphics[scale=0.29]{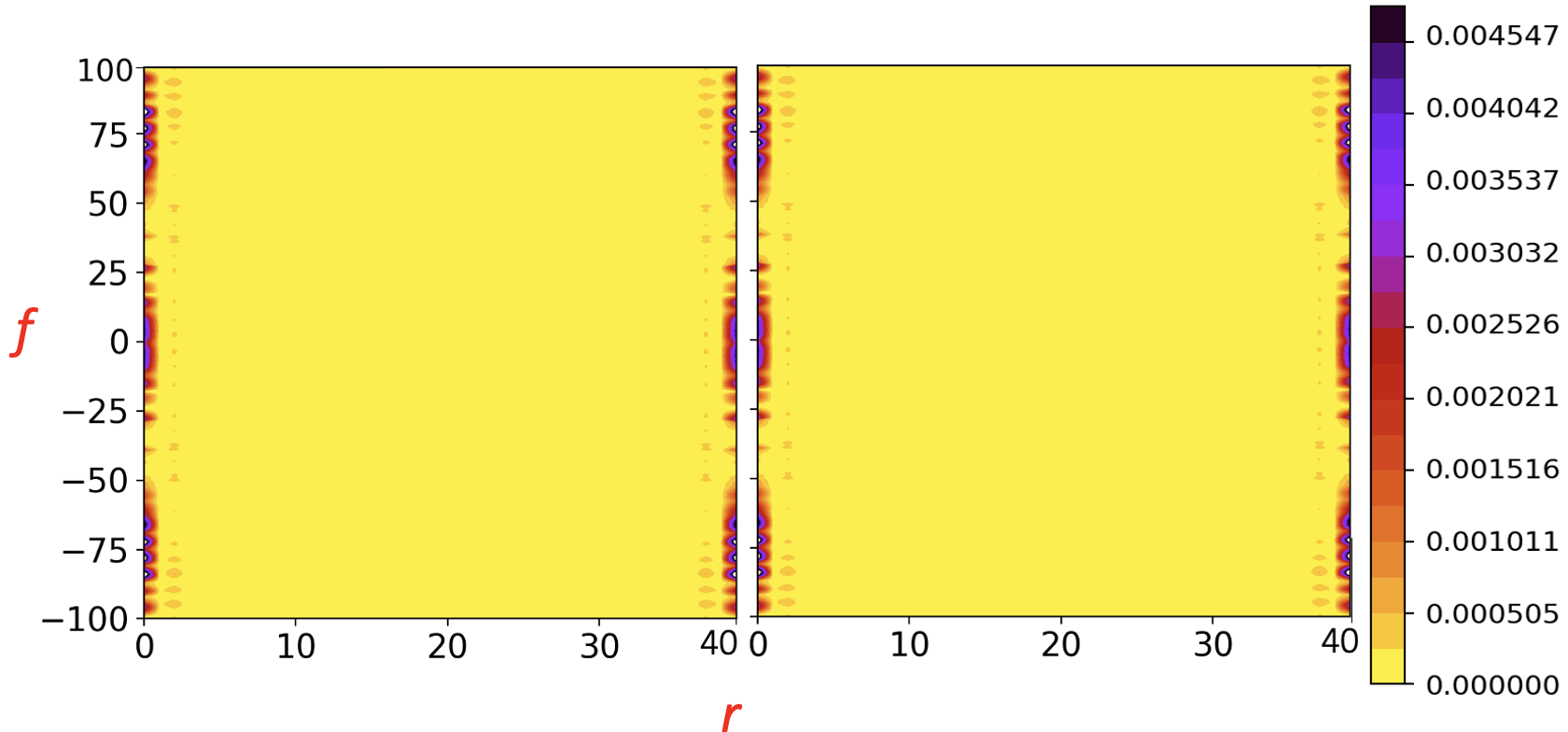}
		\caption[n-10-11]{Edge modes occurring along the real space edge for the case of $n_1,n_2=10,11$ driving frequency ratio. The contour is computed for $\psi^{\dagger}\psi_{i,j}$, for $i,j$ the real and frequency space, respectively. The states plotted are the $0$-quasi-energy states.}
		\label{fig:n-10-11}
	\end{figure}

	Since the system still possess topological order, we can plot edge modes by opening the sample boundaries. We plot the $0$-energy states in the case of broken periodic boundaries in each case. This result is shown in Fig. \ref{fig:n-10-11}. Again as in the $n_1,n_2=1,2$ we find states existing along the real space edge only.
	As noted in the case of $n_1,n_2=1,2$, we also may break translational invariance along the Fourier space. In this case the amplitude is only non-zero for certain modes along the Fourier manifold, as opposed to the modes at the very "edge". 
	The presence of mode-localized states could be an artifact of the truncation scheme, so their effect is best seen in defining and computing a physical observable in the $1$-D time dependent model.
	\section{\label{sec:Discussion}Discussion}
	\textit{Multi-Tonal Driving - Integer Multiples} There are, broadly speaking, three contrasting multi-tonal driving cases for which the Floquet formalism may be constructed. The simplest is when the frequencies are not only commensurate, but related via integer multiple. Examples for $\Omega_2/\Omega_1=n_1/n_2$ are $n_1,n_2=2,4$ or  $n_1,n_2=1,3$. In this case, one frequency may be determined in terms of the other. The relative phase of the drives plays a critical role, as explored in recent works \cite{PhysRevX.11.011057,Sandholzer_2022,Minguzzi_2022}. The formalism of Eq. \ref{ratio} is unnecessary for this driving protocol. In fact, constructing the formalism in terms of the base period may yield incorrect computation of observables. For example, treating $n_1,n_2=2,4$, with a base frequency with $n=1$ will encode an extra integer lattice spacing, yielding new non-physical twisting in the Berry curvature from $\vec{k}\rightarrow \vec{k}+d\vec{k}$. For example, the computation of the transition would yield $C=2\rightarrow C=4$, when it is physically $C=1\rightarrow C=2$.
	
	\textit{Multi-Tonal Driving - Commensurate and Incommensurate} On the other hand, there is another sort of commensurate two-tone drive where the integers cannot be uniquely expressed via an integer multiple, such as $n_1,n_2=4,5$ or  $n_1,n_2=10,11$. The Floquet theorem and topology of this case may studied with the formalism of \cite{PhysRevB.102.235143}, and the topology may be computed using the frequency space Chern marker presented here. This treatment follows from the fact that time degree of freedom should have a one-to-one correspondence with the Fourier transform to the extended space. The Floquet lattice obtains new couplings computed from the off-diagonal matrix elements of Eq. \ref{mmode_k}. This case is distinguished from the incommensurate frequency driving \cite{Martin_2017} in which each frequency yields an additional Fourier manifold. To treat the $1D$ SSH in this case would require a $3D$ computation. Additionally, two truncation schemes of the Fourier space are needed. It could be advantageous to approximate certain incommensurate ratios with nearby commensurate ones and carry out the simpler calculations presented here. 
	
	\textit{Floquet Edge States} Upon opening the "Fourier boundary" we find the zero quasi-energy modes localized to the truncation edge in the case of $n_1,n_2=1,2$, and localized to certain frequencies in the case of $n_1,n_2=10,11$. Since the Fourier boundary physically does not exist, it is simply an artifact of the theoretical mapping, these results are neglected as byproducts of the frequency truncation. The distinction between the two cases (one edge-localized and one mode-localized) could arise from the fact that the matrix size stays the same, but for higher frequencies the kinetic terms populate even further off-diagonal elements. In other words, increasing the number of Floquet modes may cause the mode-localized states to localize to the edges. While the Floquet edge is nonphysical, it is possible to break periodic boundary conditions along the Fourier manifold with small Floquet-Stark field. The consequences are best observed in this case by computing observables such as current in the original time-dependent representation.
	\section{\label{sec:Conclusions}Conclusions}
	In this paper, we have shown that commensurate multi-frequency driving formalism may be modeled using the Fourier mapping of Floquet theory to gain practical levels of engineering control. The frequency ratio may be chosen to create nontrivial couplings, which allows the Floquet formalism to mimic nontrivial static systems. We have shown that the second drive with a commensurate frequency can be included as an extra hopping term in the Fourier manifold. This approach necessitates only one Fourier manifold extension, meaning that the commensurate driving can be studied easily in $1$, and $2$-dimensional systems. Additionally, only one truncation scheme is needed in the extended space as opposed to two or more for incommensurate driving.
	
	To explore the topological properties, we have employed the real space Chern marker instead of the Berry curvature in $k$-space representation, which allows us to study the adiabatic and intermediate frequency regimes using the same framework. As such, the model could incorporate disorder and fields, more closely approximating a real world system. This treatment yields direct examination of the local fluctuations in topology resulting from interference, and of the edge states. Moreover our work suggests a new method for controlling the topological phase of an SSH sample, and appropriate choice of the frequency ratio allows for engineering of Chern numbers $C>1$. Consequentially, the amplitude proportion of the two drives may be tuned to induce topologically distinct states, meaning that these systems can be engineered to sweep through a topological phase transition. Since these topological phases are induced via the amplitude, this model hosts an experimentally appealing transition, that doesn't rely on a more complicated switching mechanism such as in the quantum wells. We have further demonstrated a computation technique to view edge states in the insulating phase, providing additional confirmation that these systems are topological insulators.
	\section{Acknowledgement}
	This work was supported by the Air Force Office of Scientific Research
	Multi-Disciplinary Research Initiative (MURI) entitled, “Cross-disciplinary
	Electronic-ionic Research Enabling Biologically Realistic Autonomous Learning
	(CEREBRAL)” under Award No. FA9550-18-1-0024 administered by Dr. Ali Sayir.
	S.W.O and W.-C.L. are grateful for the support of the summer faculty fellowship program (SFFP) sponsored by the Air-Force-Research-Lab (AFRL) while this paper is being finalized.
	\bibliography{commensuratefloquetssh.bib}
\end{document}